\documentclass[preprint2]{aastex}

\slugcomment{Submitted to AJ}

\shorttitle{}
\shortauthors{Mart\'{\i}n, Delfosse \& Guieu}

\begin{document}

\title{Spectroscopic identification of DENIS-selected brown dwarf candidates 
in the Upper Scorpius OB association\footnote{Based on observations obtained with the ESO NTT telescope.}}
\author{Eduardo L. Mart\'{\i}n\footnote{also Chercheur Associe au LAOG, Grenoble, France}}
\affil{Institute of Astronomy, University of Hawaii at Manoa,  
             2680 Woodlawn Drive, Honolulu, HI 96822, USA.\footnote{new address: Instituto de Astrof\'\i sica de Canarias, La Laguna, 38200, Spain}}
\author{Xavier Delfosse and Sylvain Guieu}
\affil{LAOG, BP 53 38041 Grenoble Cédex 9, France.}

\begin{abstract} 
We present low-resolution (R=900) optical (576.1--1,051.1~nm) 
spectroscopic observations of 40 candidate very low-mass members in the Upper Scorpius 
OB association. These objects were selected using the $I$, $J$ and $K$ photometry 
available in the DENIS database. We have derived spectral types 
and we have measured H$\alpha$ and NaI doublet (at 818.3 and 819.5~nm) equivalent widths. 
We assess the youth of the objects by comparing them to their older counterparts of 
similar spectral type in the Pleiades cluster and the field. 
Our analysis indicates that 28 of our targets are young very low-mass objects, and thus 
they are strong candidate members of the OB association. The other 12 DENIS sources are 
foreground M dwarfs or background red giants. 
Our sample of spectroscopic candidate members includes 18 objects with spectral types 
in the range M6.5 and M9, which 
are likely young brown dwarfs. We classify these candidates  
as accreting/non accreting using the scheme proposed 
by Barrado y Navascu\'es \& Mart\'\i n (2003). 
We find 5 substellar-mass candidate cluster members that are still undergoing mass accretion, 
indicating that the timescale for accretion onto brown dwarfs can be as long as 5 Myr in some cases.  
\end{abstract}


\section{Introduction}

The census of very low-mass (VLM) stars and 
brown dwarfs (BDs) is very incomplete, even in the immediate solar neighborhood, 
and in the nearest open clusters, star-forming regions and OB associations. 
Considerable progress has been achieved in the last few years, leading to the identification 
of hundreds of VLM stars and BDs and to the development of two new 
spectral classes, namely L and T, for ultracool dwarfs (Mart\'\i n et al$.$ 1997; Kirkpatrick et al$.$ 1999; 
Jones \& Steele 2001; Burgasser et al$.$ 2002; Geballe et al$.$ 2002). 
This huge observational effort has proved that VLM stars and BDs constitute a 
numerous population in the galactic disk. They are being incorporated in our current paradigm of 
star and planet formation 
(Boss 2002; Bate et al$.$ 2002; Delgado-Donate et al$.$ 2003; Kroupa et al$.$ 2001, 2003; 
Sterzik \& Durisen 2002) and stellar evolution (D'Antona 
\& Mazzitelli 1994; Baraffe et al$.$ 1998, 2002; Wuchterl \& Tscharnuter 2003).  

One of the most successful methods to identify VLM objects has been to obtain deep enough 
multicolor imaging of the classical young associations and clusters in the solar vicinity, 
and follow-up spectroscopy of the candidates that occupy the expected position of VLM 
members in the color-magnitude diagram. In the Pleiades open cluster, where the BD population is well  
studied, the success rate of imaging plus low-resolution spectroscopic characterization is 
about 90\% (Mart\'\ in et al$.$ 1996, 2000; Moraux et al$.$ 2001). 
In younger regions, the success rate could be even higher because it is easier 
to apply gravity sensitive spectroscopic indicators due to the lower surface gravity of 
younger VLM objects. 

Upper Scorpius (hereafter USco for brevity) is the youngest subgroup (age$\sim$5~Myr, de Geus 
et al$.$ 1989)  of the nearest OB association (Scorpius-Centaurus). The average distance 
to the association 
is 145~pc, but there is a large scatter in the Hipparcos parallaxes of individual members 
(de Geus et al$.$ 1989). The low-mass stellar population of USco has been 
studied by Walter et al$.$ (1994), Mart\'\i n (1998), and Preibisch et al$.$ (1998, 2001). 
The VLM population was first revealed by Ardila et al$.$ (2000). 

In this paper we present the spectroscopic characterization of 36 new candidate VLM members in the 
USco association. 3 of our targets (cross identifications are given in Table~1) 
were already found by Ardila et al. (2000). One of them, namely DENIS161007.5-181056.4,  
was already included in the survey of Preibisch et al. (2001).  
Our candidates were identified in the database of the DEep Near-Infrared 
Survey (DENIS) using $I$, $J$ and $K$ photometry (Epchtein et al$.$ 1994). DENIS 
has already been mined to identify nearby field VLM stars and BDs (Delfosse et al$.$ 
1997, 1999; PhanBao et al$.$ 2001, 2003). A preliminary report of the results of this 
survey can be found in Delfosse et al$.$ (2003). 
A paper dedicated to the BD initial mass function (IMF) in a rich star formation
region -such as an OB association-  and the distribution of VLM USco members, 
will be forthcoming (Delfosse et al$.$ in prep). The comparison of the BD
IMF for star formation regions with  different densities (from OB
associations to T Tauri associations) will be a stepping stone to 
understand BD formation. The present paper is focused on the 
spectroscopic follow up and identification of the DENIS BD  
candidates in the USco association.   

In section 2, we present the observations and data reductions. 
Section 3 deals with the spectroscopic measurements and data analysis. 
In section 4 we assess the membership of the candidates in USco. 
Section 5 consists of a discussion of the properties of the objects.


\section{Sample selection, observations and data reduction}

Our sample of 40 VLM candidate members in USco was selected from
the DENIS database. The DENIS 100\% completeness limit is I=18.0,
which for the NextGen models (Baraffe et al. 1998)  
translates into a 20~M$_{\rm Jup}$ brown dwarfs at the age ($\sim$~5Myr)   
and average distance ($\sim$~145pc) of the association. Our search was
made in 60 square degrees of the association, centered at the galactic
coordinate $l=347^{o}$, $b=25^{o}$.  The already
known members of the association are mainly located in a band 2~mag
higher than the main-sequence in the H-R diagram. 
Preibisch et al. (2001) find that only $\sim$10\% of
USco stars are outside such band. In Figure~1  
 we show the color-magnitude diagram for our DENIS survey in
USco. The NextGen 5Myr theoreticaly isochrones for distances of 120~pc 
and 200~pc mark the band of expected location for the
association members. Our search is focused on the red part of this
band. After artifact rejection, 104 objects were retained as candidate with
color redder than $I-J=2.3$. Nine of them were already detected by Ardila et al. (2000), 
and one was reported by Preibisch et al. (2001), all the other are new. 

The candidates of Ardila et al. (2000) detected in the 60 square 
degrees of our DENIS survey  are
marked in Figure~1. 15\% of them are outside the band. Our 
search is not complete because it lacks over-luminous or
sub-luminous members (e.g. binary or obscured objects), but we are available
to detect 80-90\% of Ardila's candidate members. We refer the reader to Delfosse et
al. (in preparation) for a detailed analysis, dedicated to
determine luminosity and mass functions of the brown dwarf sequence in
the association. 

Optical spectroscopy was obtained for 40 objects out of our 104 candidates. The
spectroscopic target list is constituted by the redder candidates (all
except two for $I-J>3.0$) and by candidate regularly spaced out on the rest
of the sequence to estimate the association membership fractions of our
candidates in function of their localization in the color-magnitude diagram.
The spectroscopic observations were obtained in service mode at the ESO NTT telescope 
using the EMMI spectrograph on the nights of 13-16 June 2002 and 7-8 August 2002. 
Exposure times for each target are given in Table~1. 
The CCD frames were trimmed, dark subtracted, and flat fielded before extracting 
one dimensional spectra. The spectra were wavelength calibrated using an arc 
lamp spectrum, 
and flux calibrated using a standard star spectrum. 
All reduction operations were carried out in the IRAF environment. 
The wavelength coverage 
was 576.1--1,051.1~nm at a dispersion of 0.177 nm/pix and a 
resolving power of R=900. In Figure~2 we display a representative subset of our 
final spectra. The main spectral features have been identified following 
Kirkpatrick et al$.$ (1991) and Mart\'\i n et al$.$ (1996).

\section{Analysis}

In Table~2 we provide the measurements obtained from the spectra of our targets. 
We have measured W(H$\alpha$) and W(NaI) at 818.3 and 919.0~nm using the IRAF task splot. The 
pseudocontinuum was placed by visual inspection of the spectra. The measurement 
errors are dominated by uncertainties in the location of the pseudocontinuum. 
Spectroscopic indices of the strength of TiO and VO bands, and the slope 
of the pseudocontinuum (PC3), were also evaluated using the 
definitions in Mart\'\i n et al$.$ (1999). We used the PC3-spectral type calibration 
provided by these authors to obtain spectral types for our objects. We checked that the 
TiO and VO values were consistent with the spectral type determined from the PC3 index. 
We do not find evidence for any significant reddening in our sample. We estimate 
that Av$<$1 for most of our objects. Ardila et al$.$ (2000) found only two objects with Av$>$1 
among 23 VLM USco candidates. Our dataset is not as sensitive to reddening as that of 
Ardila et al$.$ because we lack the $R-I$ color. 

\subsection{Cluster membership assessment} 

We assess the membership of the candidates in USco on the basis of their location in 
the magnitude-spectral type and W(NaI)-spectral type diagrams. 

In Figure~3 we plot our targets in a spectral type - magnitude diagram. Note that six  
objects are located outside the cluster sequence. They have spectral types that are too 
early for their $I$ magnitude. We consider them as likely non-members. Five of them 
do not present H$\alpha$ in emission. They are probably background red giants. 
We discard them from further analysis because they are not interesting for the 
purposes of this paper. 

The NaI doublet at 818.3 and 819.5~nm is a  well known gravity indicator for late-M spectral types 
(Kirkpartrick et al$.$ 1991; Mart\'{\i}n et al$.$ 1996; Brice\~no et al$.$ 1998). It becomes 
weaker for decreasing surface gravity. In Figure~4 we plot our W(NaI) values versus spectral type 
and we compare with the field dwarfs measured by  Mart\'{\i}n et al$.$ (1996). 
VLM USco candidates that W(NaI) similar to those of field dwarfs are considered likely foreground 
cool dwarfs. The USco members are expected to be young, and hence they should have larger 
radius, lower surface gravity and weaker W(NaI) than their field counterparts of the same 
spectral type. The USco candidates with W(NaI) weaker than the field objects are retained as 
likely cluster members.  

H$\alpha$ emission is normally seen among young VLM stars and BDs. It can originate 
from chromospheric activity or from disk accretion onto the central object. 
We expect that the young USco VLM members have H$\alpha$ emission because they are 
chromospherically active and may still have some leftover accretion activity. 
We do not use H$\alpha$ emission as a membership criterion, but we note that all 
the 28 likely members listed in Table~2 clearly have H$\alpha$ in emission.

\subsection{Mass accretion}

In order to identify classical T Tauri stars and substellar analogs using only low-resolution 
optical spectroscopy, 
Barrado y Navascu\'es \& Mart\'\i n (2003) have proposed a W(H$\alpha$) boundary 
line that depends on the spectral type, and it is 
based on the chromospheric saturation limit at Log \{L(H$\alpha$)/L(bol)\}=$-$3.3, 
observed in young open clusters. 
Objects with  W(H$\alpha$) larger than the boundary are likely accretors. 
In Figure~5 we display the  W(H$\alpha$) of our 
USco objects as a function of spectral type. We also show the Barrado y Navascu\'es \& Mart\'\i n 
boundary line. Six objects from our sample, and one from Ardila et al$.$'s sample, are located 
above the line. Hence, they are likely accretors. Five of the accretors have spectral types 
later than M6, so they are probably substellar\footnote{
The customary spectral type boundary between stars and BDs 
for ages younger than about 120~Myr is at M6 (Mart\'{\i}n et al$.$ 1996;  
Luhman et al$.$ 1998.}. 

Barrado y Navascu\'es \& Mart\'\i n (2003) have combined our results with the 
literature to estimate a frequency of accretors of 14.4$\pm$5.7\% for the spectral type 
range K3-M5.5, while for the spectral type range M5.5-M9 the frequency is 16.3$\pm$6.2\% . 
Hence, there is no evidence for a strong dependence of the number of accreting objects 
with primary mass in the USco low-mass population. Other studies based on infrared 
color excesses have reached similar conclusions (Haisch et al$.$ 2001; Muench et al$.$ 2001;  
Natta et al$.$ 2002; Jayawardhana et al$.$ 2003; Liu et al$.$ 2003).  
Our sample of USco likely substellar accretors  
deserves further follow-up observations to determine their spectral energy distribution and 
mass accretion rates.   

\section{Discussion}

In this paper we have presented new evidence for a numerous VLM population in the USco 
OB association. We refer the reader to Delfosse et al. (in preparation) for a discussion of the VLM 
luminosity and mass functions in USco. 

We have identified 28 objects with spectral types in the range M5.5-M9 
as likely VLM members on the basis of their spectroscopic characteristics. They 
are located in the cluster sequence in the magnitude-spectral type diagram, they have 
relatively weak NaI lines, and they show H$\alpha$ in emission. Five of them 
meet the Barrado y Navascu\'es \& Mart\'\i n criterion for substellar classical T Tauri analogs. 
Since these accreting BDs are likely members of the OB association, their age is 5~Myr, implying 
that accretion onto BDs can last a significant amount of time. 

Together with the VLM candidates presented by Ardila et al. (2000), and the follow-up spectroscopic 
observations of 12 of them (none of which overlap with our sample) 
by Mohanty et al$.$ (2003), our study brings the number of spectroscopically 
identified USco VLM candidate members to 40. The membership of the objects 
needs to be confirmed with follow-up studies of proper motion and radial velocities. 
This is now a significant sample for systematic studies of the properties of 
young VLM objects, such as accretion, binarity and rotation. 
Due to its relative proximity, young age, and richness, 
USco is one of the most important regions for studies of substellar-mass formation and evolution. 
To compare the IMF in such a rich star formation
region with the one of less dense regions (as the T Tauri associations) will 
be very useful to constrain the sensitivity of BD formation to environmental conditions.


\begin{acknowledgements}
EM thanks the staff of the LAOG for their hospitality during his visit. 
We thank Jana Pittichova for helping with data reduction and  
Herve Bouy for useful comments about some of the targets. 
Partial funding was provided by the National Aeronautics and Space 
Administration (NASA) grant NAG5-9992 and 
National Science Foundation (NSF) grant AST-0205862. 

\end{acknowledgements}


\clearpage

\begin{table}
\caption[]{USco DENIS candidate members}
\begin{tabular}{lrrrrr}
\hline
Name & $I$ & $J$ & $K$ & t$_{\rm exp}$ (s) & Notes \\   

\hline
\hline
DENIS-PJ151450.1-225435.3 &  17.08$\pm$0.09 & 13.96$\pm$0.08 & 12.86$\pm$0.17 & 2x510 & \\
DENIS-PJ151713.7-200238.9 &  15.40$\pm$0.10 & 13.07$\pm$0.07 & 12.07$\pm$0.10 & 300   & \\
DENIS-PJ153338.7-201856.0 &  16.70$\pm$0.09 & 14.13$\pm$0.10 & 13.03$\pm$0.12 & 2x300 & \\
DENIS-PJ155504.9-204258.3 &  17.90$\pm$0.18 & 14.75$\pm$0.15 & 13.48$\pm$0.22 & 2x600 & \\
DENIS-PJ155556.0-204518.5 &  16.20$\pm$0.07 & 13.43$\pm$0.10 & 11.94$\pm$0.11 & 2x300 & \\
DENIS-PJ155601.0-233808.1 &  16.32$\pm$0.09 & 13.96$\pm$0.11 & 12.85$\pm$0.14 & 2x300 & UScoCTIO113 \\
DENIS-PJ155605.0-210646.4 &  16.85$\pm$0.09 & 14.11$\pm$0.12 & 12.74$\pm$0.13 & 2x450 & \\
DENIS-PJ160019.5-225628.4 &  17.45$\pm$0.12 & 14.67$\pm$0.14 & 13.41$\pm$0.18 & 2x510 & UScoCTIO131 \\
DENIS-PJ160334.7-182930.4 &  14.84$\pm$0.07 & 12.47$\pm$0.07 & 11.43$\pm$0.09 & 300   & \\
DENIS-PJ160440.8-193652.8 &  16.09$\pm$0.06 & 13.52$\pm$0.10 & 12.40$\pm$0.11 & 2x300 & \\
DENIS-PJ160455.8-230743.8 &  16.45$\pm$0.08 & 13.88$\pm$0.11 & 12.71$\pm$0.13 & 2x300 & UScoCTIO117  \\
DENIS-PJ160514.0-240652.6 &  15.22$\pm$0.05 & 12.78$\pm$0.08 & 11.94$\pm$0.10 &  300  & \\
DENIS-PJ160603.9-205644.6 &  16.38$\pm$0.09 & 13.54$\pm$0.08 & 12.37$\pm$0.10 & 2x300 & \\
DENIS-PJ160809.0-272748.0 &  16.46$\pm$0.08 & 14.11$\pm$0.13 & 13.30$\pm$0.17 & 2x300 & \\
DENIS-PJ160951.1-272242.2 &  15.68$\pm$0.06 & 13.33$\pm$0.09 & 12.33$\pm$0.11 & 2x300 & \\
DENIS-PJ160958.5-234518.6 &  15.04$\pm$0.04 & 12.66$\pm$0.08 & 11.49$\pm$0.08 & 300   & \\
DENIS-PJ161005.4-191936.0 &  17.16$\pm$0.12 & 14.21$\pm$0.12 & 12.80$\pm$0.14 & 2x500 & \\
DENIS-PJ161006.0-212744.6 &  18.11$\pm$0.20 & 14.74$\pm$0.15 & 13.59$\pm$0.22 & 2x600 & \\
DENIS-PJ161007.5-181056.4 &  15.11$\pm$0.05 & 12.80$\pm$0.08 & 11.70$\pm$0.09 & 300   & \\
DENIS-PJ161030.1-231516.7 &  17.04$\pm$0.11 & 14.37$\pm$0.13 & 13.19$\pm$0.17 & 2x450 & \\
DENIS-PJ161050.0-221251.8 &  15.05$\pm$0.04 & 12.68$\pm$0.08 & 11.64$\pm$0.10 & 300   & \\
DENIS-PJ161103.6-242642.9 &  17.80$\pm$0.16 & 14.68$\pm$0.15 & 13.34$\pm$0.17 & 2x600 & \\
DENIS-PJ161420.6-274549.6 &  17.42$\pm$0.11 & 14.59$\pm$0.11 & 13.55$\pm$0.22 & 2x510 & \\
DENIS-PJ161452.6-201713.2 &  18.56$\pm$0.20 & 15.33$\pm$0.15 &                & 3x900 & \\
DENIS-PJ161624.0-240830.2 &  15.51$\pm$0.11 & 13.12$\pm$0.06 & 12.14$\pm$0.11 & 300   & \\
DENIS-PJ161632.2-220520.2 &  16.06$\pm$0.12 & 13.71$\pm$0.08 & 12.66$\pm$0.13 & 2x300 & \\
DENIS-PJ161758.1-265034.0 &  16.76$\pm$0.08 & 14.25$\pm$0.10 & 12.82$\pm$0.16 & 2x450 & \\
DENIS-PJ161816.2-261908.1 &  14.60$\pm$0.03 & 12.20$\pm$0.07 & 10.90$\pm$0.06 & 300   & \\
DENIS-PJ161820.5-260007.8 &  16.91$\pm$0.09 & 14.28$\pm$0.11 & 12.96$\pm$0.14 & 2x450 & \\
DENIS-PJ161833.2-251750.4 &  15.03$\pm$0.04 & 12.51$\pm$0.07 & 11.22$\pm$0.06 & 300   & \\
DENIS-PJ161840.8-220948.1 &  16.36$\pm$0.07 & 13.79$\pm$0.09 & 12.84$\pm$0.13 & 2x300 & \\
DENIS-PJ161855.1-260035.2 &  18.09$\pm$0.18 & 15.06$\pm$0.14 &                & 2x600 & \\
DENIS-PJ161903.4-234408.8 &  17.00$\pm$0.09 & 14.18$\pm$0.10 & 12.74$\pm$0.13 & 2x450 & \\
DENIS-PJ161916.5-234722.9 &  17.98$\pm$0.15 & 14.83$\pm$0.13 & 13.70$\pm$0.22 & 2x450 & \\
DENIS-PJ161926.4-241244.5 &  15.96$\pm$0.05 & 13.56$\pm$0.09 & 12.29$\pm$0.10 & 2x300 & \\
DENIS-PJ161929.9-244047.1 &  17.32$\pm$0.10 & 14.21$\pm$0.10 & 12.86$\pm$0.13 & 2x510 & \\
DENIS-PJ161939.8-214535.1 &  15.64$\pm$0.04 & 13.21$\pm$0.08 & 12.07$\pm$0.09 & 2x300 & \\
DENIS-PJ162037.8-242757.8 &  17.56$\pm$0.13 & 14.57$\pm$0.12 & 13.08$\pm$0.16 & 2x510 & \\
DENIS-PJ162041.5-242549.0 &  17.52$\pm$0.13 & 14.42$\pm$0.11 & 13.00$\pm$0.15 & 2x510 & \\
DENIS-PJ162058.0-200846.1 &  17.67$\pm$0.14 & 14.23$\pm$0.11 & 11.98$\pm$0.10 & 2x510 & \\
\hline
\end{tabular}
\end{table}

\clearpage

\begin{table}
\caption[]{Spectroscopic measurements and membership assessment}
\begin{tabular}{lrrrrrrr}
\hline
Name                 & W(H$\alpha$)   & PC3           & TiO & VO & W(NaI)        & SpT & Member? \\   
                     &  (\AA)         &               &     &    &  (\AA)        &     &         \\
\hline
\hline
DENIS-PJ151450.1-225435.3 &  $>$-6.5  & 2.40$\pm$0.03 & 3.4 &  2.6 & 6.8$\pm$0.6 & M9.5 & No \\
DENIS-PJ151713.7-200238.9 &  -5$\pm$1 & 1.48$\pm$0.07 & 3.4 &  2.4 & 7.2$\pm$0.7 & M6   & No \\
DENIS-PJ153338.7-201856.0 &  -6$\pm$2 & 1.84$\pm$0.04 & 3.9 &  2.6 & 8.2$\pm$0.5 & M8   & No \\
DENIS-PJ155504.9-204258.3 & -14$\pm$5 & 1.95$\pm$0.07 & 4.5 &  2.7 & 7.0$\pm$1.0 & M8.5 & No \\
DENIS-PJ155556.0-204518.5 & -11$\pm$3 & 1.50$\pm$0.03 & 2.9 &  2.5 & 3.6$\pm$0.3 & M6.5 & Yes \\
DENIS-PJ155601.0-233808.1 & -20$\pm$3 & 1.52$\pm$0.01 & 4.0 &  2.7 & 5.2$\pm$0.2 & M6.5 & Yes \\ 
DENIS-PJ155605.0-210646.4 & -20$\pm$2 & 1.64$\pm$0.05 & 4.0 &  2.8 & 4.7$\pm$0.3 & M7   & Yes \\
DENIS-PJ160019.5-225628.4 &  -5$\pm$3 & 1.81$\pm$0.04 & 4.3 &  2.8 & 5.2$\pm$0.8 & M8   & Yes \\ 
DENIS-PJ160334.7-182930.4 & -19$\pm$1 & 1.37$\pm$0.04 & 3.1 &  2.4 & 3.5$\pm$0.3 & M5.5 & Yes \\
DENIS-PJ160440.8-193652.8 & -16$\pm$2 & 1.52$\pm$0.01 & 3.7 &  2.6 & 4.9$\pm$0.3 & M6.5 & Yes \\
DENIS-PJ160455.8-230743.8 & -25$\pm$3 & 1.56$\pm$0.01 & 3.9 &  2.7 & 5.3$\pm$0.4 & M6.5 & Yes \\ 
DENIS-PJ160514.0-240652.6 & -24$\pm$1 & 1.42$\pm$0.03 & 3.6 &  2.6 & 3.4$\pm$0.3 & M6   & Yes \\
DENIS-PJ160603.9-205644.6 & -105$\pm$12 & 1.68$\pm$0.04 & 3.7 & 2.7 & 4.2$\pm$0.3 & M7.5 & Yes \\
DENIS-PJ160809.0-272748.0 &  -6$\pm$1 & 1.32$\pm$0.01 & 2.6 & 2.3  & 5.1$\pm$0.2 & M5   & No \\
DENIS-PJ160951.1-272242.2 & -22$\pm$2 & 1.44$\pm$0.02 & 3.6 & 2.6  & 4.5$\pm$0.1 & M6   & Yes \\
DENIS-PJ160958.5-234518.6 & -20$\pm$1 & 1.51$\pm$0.02 & 3.6 & 2.6  & 4.5$\pm$0.2 & M6.5 & Yes \\
DENIS-PJ161005.4-191936.0 & -48$\pm$4 & 1.65$\pm$0.01 & 3.5 & 2.7  & 3.2$\pm$0.1 & M7   & Yes \\
DENIS-PJ161006.0-212744.6 & -17$\pm$3 & 2.02$\pm$0.01 & 4.3 & 2.9  & 3.8$\pm$0.1 & M8.5 & Yes \\
DENIS-PJ161007.5-181056.4 & -19$\pm$1 & 1.46$\pm$0.01 & 3.6 & 2.5  & 4.1$\pm$0.2 & M6   & Yes \\
DENIS-PJ161030.1-231516.7 & -12$\pm$4 & 1.74$\pm$0.04 & 4.2 & 2.7  & 7.6$\pm$0.7 & M7.5 & No \\
DENIS-PJ161050.0-221251.8 & -23$\pm$1 & 1.39$\pm$0.01 & 3.3 & 2.5  & 4.5$\pm$0.1 & M5.5 & Yes \\
DENIS-PJ161103.6-242642.9 & -28$\pm$3 & 2.07$\pm$0.05 & 4.3 & 2.9  & 3.8$\pm$0.2 & M9   & Yes \\
DENIS-PJ161420.6-274549.6 & -26$\pm$2 & 1.76$\pm$0.01 & 4.0 & 2.7  & 6.3$\pm$0.9 & M7.5 & No \\
DENIS-PJ161452.6-201713.2 & -51$\pm$5 & 2.10$\pm$0.05 & 4.3 & 3.0  & 4.4$\pm$0.5 & M9   & Yes \\
DENIS-PJ161624.0-240830.2 & -13$\pm$1 & 1.35$\pm$0.03 & 3.1 & 2.4  & 4.2$\pm$0.2 & M5.5 & Yes \\
DENIS-PJ161632.2-220520.2 & -12$\pm$1 & 1.43$\pm$0.02 & 3.1 & 2.4  & 4.0$\pm$0.2 & M6   & Yes \\
DENIS-PJ161758.1-265034.0 & $>$-3     & 1.51$\pm$0.04 & 2.4 & 2.3  & 3.5$\pm$0.3 & M5   & No  \\
DENIS-PJ161816.2-261908.1 & -9$\pm$1  & 1.38$\pm$0.03 & 2.6 & 2.4  & 1.4$\pm$0.1 & M5.5 & Yes \\
DENIS-PJ161820.5-260007.8 &  $>$-2    & 1.67$\pm$0.10 & 1.9 & 2.1  & 3.3$\pm$1.0 & M3.5 & No \\
DENIS-PJ161833.2-251750.4 & -17$\pm$1 & 1.49$\pm$0.10 & 2.6 & 2.4  & 3.0$\pm$0.2 & M6   & Yes \\
DENIS-PJ161840.8-220948.1 & -11$\pm$2 & 1.63$\pm$0.02 & 3.6 & 2.6  & 4.2$\pm$0.9 & M7   & Yes \\
DENIS-PJ161855.1-260035.2 &  $>$-1    & 1.30$\pm$0.02 & 2.4 & 2.2  & 3.3$\pm$1.0 & M3   & No \\
DENIS-PJ161903.4-234408.8 & -15$\pm$6 & 1.57$\pm$0.01 & 2.9 & 2.5  & 3.5$\pm$0.2 & M6.5 & Yes \\
DENIS-PJ161916.5-234722.9 & -125$\pm$10 & 1.89$\pm$1.12 & 2.7 & 2.5 & 2.0$\pm$0.4 & M8 & Yes \\
DENIS-PJ161926.4-241244.5 & -14$\pm$1 & 1.46$\pm$0.01 & 3.1 & 2.5  & 3.6$\pm$0.2 & M6 & Yes \\
DENIS-PJ161929.9-244047.1 & -65$\pm$15 & 1.80$\pm$0.02 & 3.9 & 2.7 & 3.6$\pm$0.2 & M8 & Yes \\
DENIS-PJ161939.8-214535.1 & -61$\pm$7 & 1.60$\pm$0.03 & 3.5 & 2.4 & 5.2$\pm$0.8 & M7 & Yes \\
DENIS-PJ162037.8-242757.8 &  $>$-2    & 1.63$\pm$0.03 & 2.4 & 2.3 & 3.8$\pm$0.5 & M4 & No \\
DENIS-PJ162041.5-242549.0 & -11$\pm$8 & 1.74$\pm$0.01 & 3.6 & 2.8 & 3.8$\pm$0.5 & M7.5 & Yes \\
DENIS-PJ162058.0-200846.1 &  $>$-3    & 1.91$\pm$0.05 & 1.8 & 2.0 & 3.8$\pm$0.5 & M3  & No \\
\hline
\end{tabular}
\end{table}

\setcounter{figure}{0}
   \begin{figure*}
   \centering
   \includegraphics[width=14.0cm]{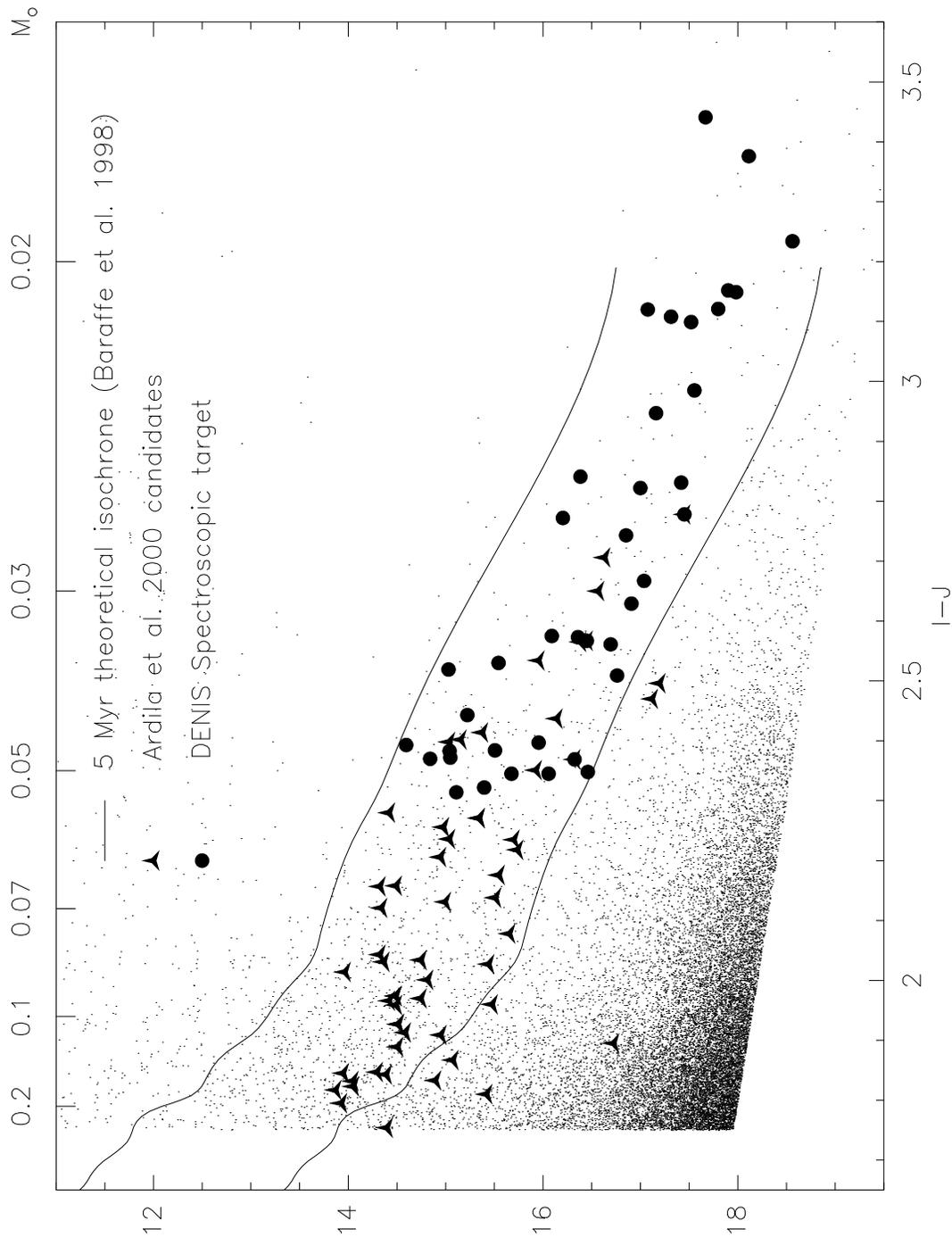}
\caption{$I$ vs $I-J$ color-magnitude diagram for our targets for 
spectroscopic follow-up detected
  by DENIS in 60 squares degrees of the Upper Scorpius OB association. The
  two solid lines delimit the area of candidate members, based on the
  Baraffe et al. 1998 theoretical isochrone for 5 Myr and distances of 
120~pc and 200~pc. The dots are objects in the
  DENIS database (inluding artifacts), and the empty circles are DENIS spectroscopic
  targets discussed in the present paper. The triangles are brown 
  dwarfs that were also previously identified by Ardila et al$.$ (2000). The
  mass labels on the top 
  axis are deduced from $I-J$ using the Baraffe et al. (1998) models. 
 }
         \label{}
   \end{figure*}

\clearpage
\setcounter{figure}{1}
   \begin{figure*}
   \centering
   \includegraphics[width=14.0cm]{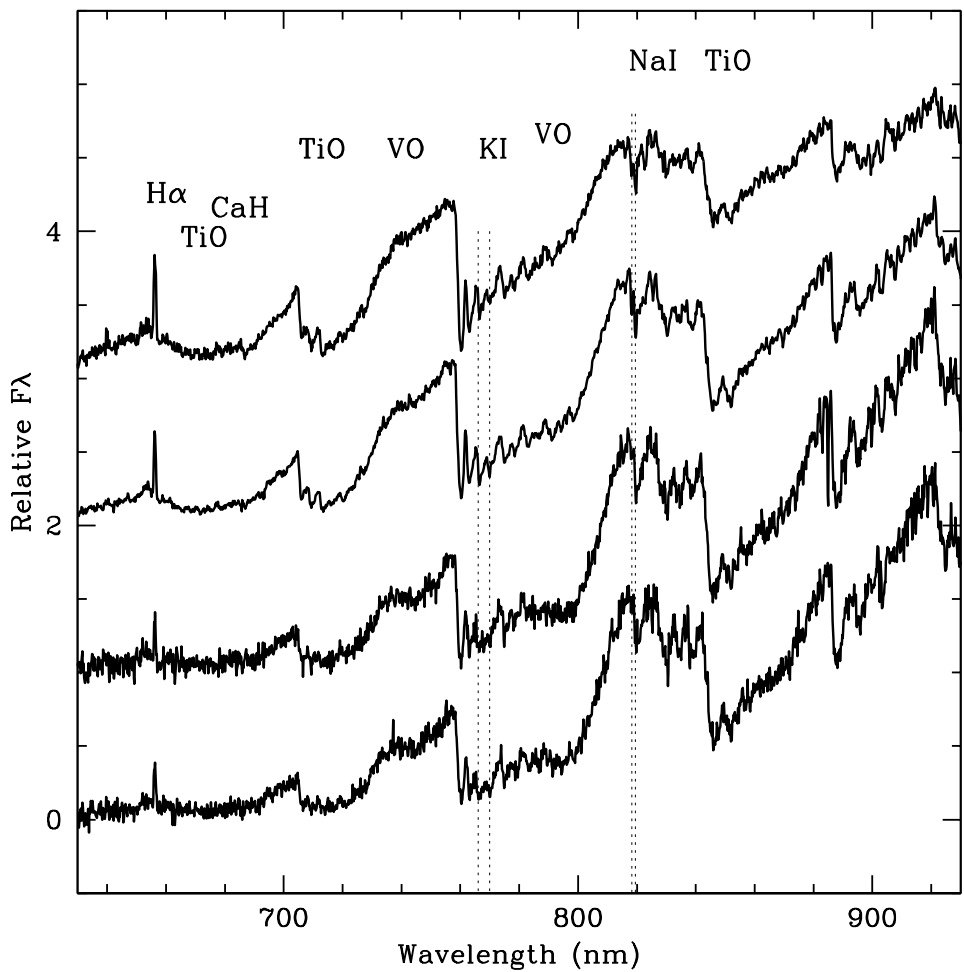}
\caption{Final spectra for a representative subset of our targets. From top to bottom: 
DENIS-PJ160334.7-182930.4 (M5.5), DENIS-PJ160951.1-272242.2 (M6.5), 
DENIS-PJ161006.0-212744.6 (M8.5), and DENIS-PJ161103.6-242642.9 (M9).}
         \label{}
   \end{figure*}

\clearpage
\setcounter{figure}{2}

   \begin{figure*}
   \centering
   \includegraphics[width=14.0cm]{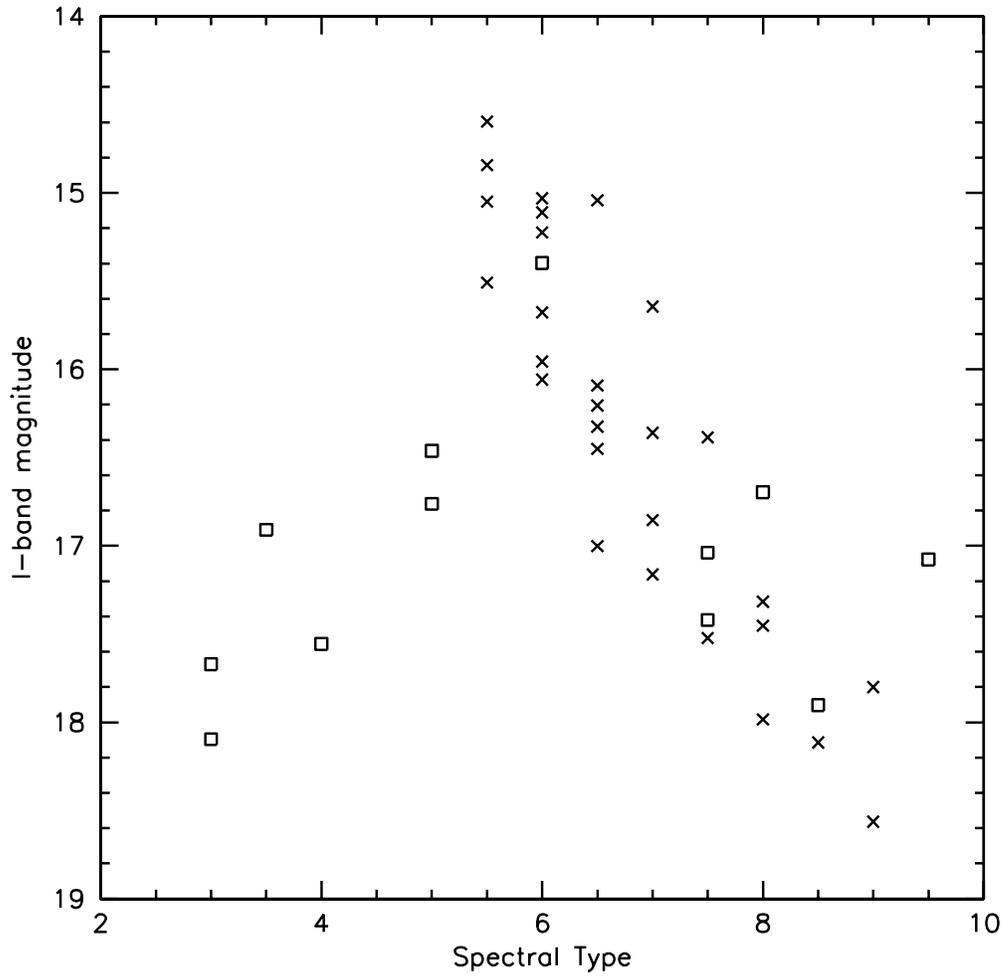}
\caption{Spectral type - magnitude diagram for our program objects. Crosses denote 
likely USco members on the basis of their spectral type, H$\alpha$ equivalent width and 
NaI equivalent width. Empty squares denote probable non-members.}
         \label{}
   \end{figure*}


\clearpage
\setcounter{figure}{3}

   \begin{figure*}
   \centering
   \includegraphics[width=14.0cm]{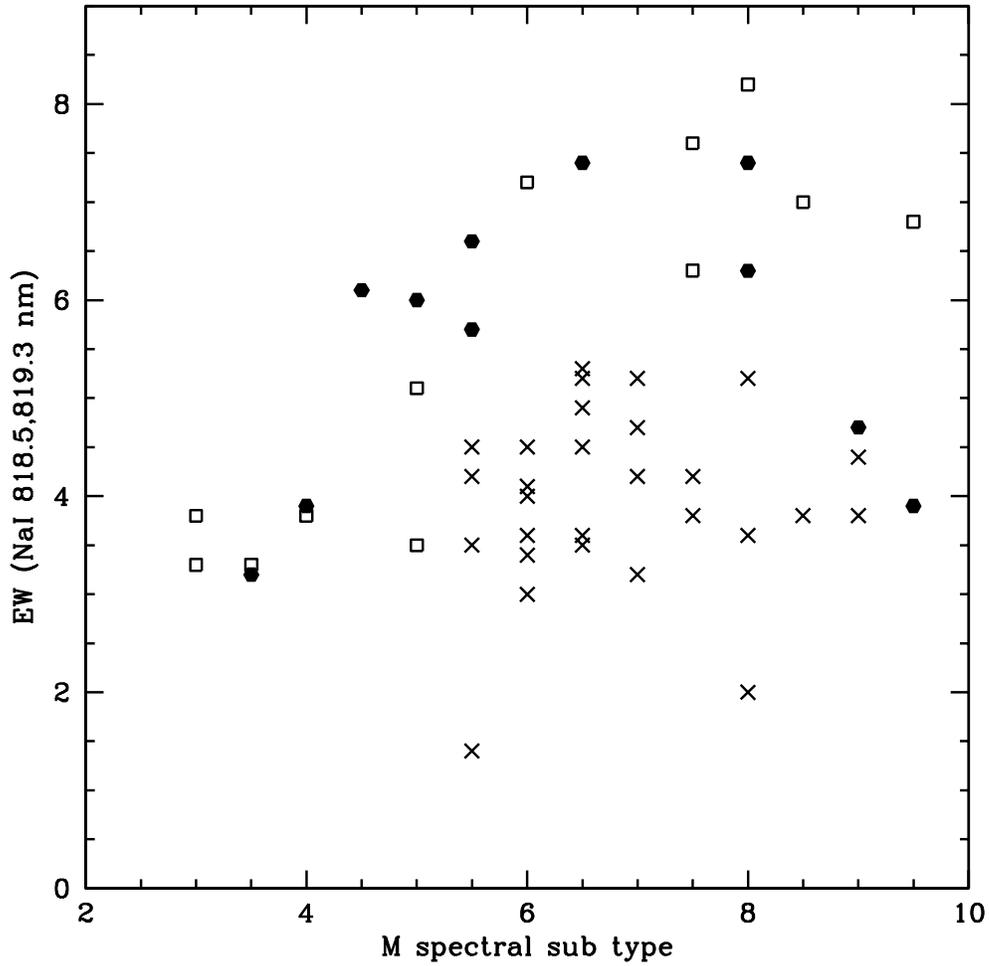}
\caption{Spectral type - W(NaI) diagram for our program objects. Crosses denote 
likely USco members. Empty squares denote probable non-members. Filled hexagons 
represent field dwarfs from Mart\'\i n et al$.$ 1996. }
         \label{}
   \end{figure*}


\clearpage
\setcounter{figure}{4}

   \begin{figure*}
   \centering
   \includegraphics[width=14.0cm]{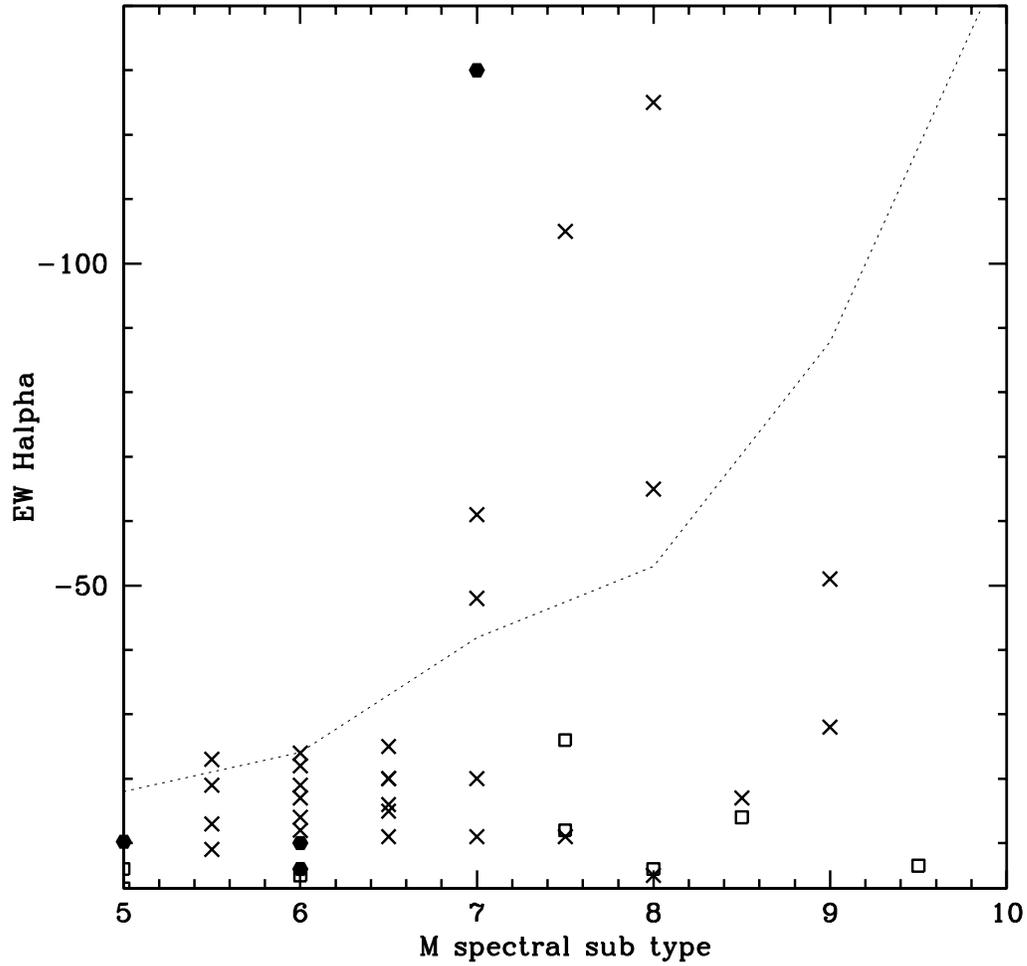}
\caption{Spectral type - W(H$\alpha$) diagram for our program objects. Crosses denote 
likely USco members. Empty squares denote probable non-members. Filled hexagons 
represent USco candidates from Ardila et al$.$ 2000. The dashed line marks the boundary 
between accreting and non-accreting objects according to the saturation criterion 
defined by Barrado y Navascu\'es \& Mart\'\i n (2003).}
         \label{}
   \end{figure*}

\end{document}